\documentclass[twoside]{elsart}
\usepackage{epsfig}
\usepackage{array}
\usepackage{amssymb,amsbsy,amsmath}
\newcommand{\skipit}[1]{}
\newcommand{\figref}[1]{fig.~\protect\ref{#1}}
\newcommand{\xref}[1]{\protect\ref{#1}}

\newcommand{\fmref}[1]{(\protect\ref{#1})}
\newlength{\CaptionWidth}
\newcommand{\mycaption}[2]{%
\begin{center}\begin{minipage}{\CaptionWidth}%
\caption[]{#1}\label{#2}%
\end{minipage}\end{center}%
}%
\def\V0{\stackrel{\circ}{V}}
\def\v0{\stackrel{\circ}{v}}
\def\half{{\frac{1}{2}}\;}

\newcommand{\Th}{{\mbox{\scriptsize Th}}}

\newcommand{\fm}{\mbox{\,fm}}
\newcommand{\MeV}{\mbox{\,MeV}}
\newcommand{\dint}{\mbox{d}}

\newcommand{\Tr}{\mbox{Tr}}
\newcommand{\tr}{\mbox{tr}}
\newcommand{\op}[1]{\ensuremath{%
    \fontdimen12\textfont3=2pt\fontdimen12\scriptfont3=1.4pt%
    \!\null\mathop{\vphantom{#1}\smash{#1}}\limits_{\sim}\null\!}}
\newcommand{\Operator}[1]{\smash{\raisebox{-1.1ex}{
$\!\!\stackrel{\displaystyle #1}{\sim}$}}}

\newcommand{\OpRerg}
           {\Operator{R}_{{\vphantom{A}}^{\mbox{\scriptsize erg}}}}
\newcommand{\EnsembleMean}[1]{\langle\langle \; {#1}\; 
            \rangle\rangle_{\big|\;\scriptstyle{T}} \;}

\newcommand{\TimeAv}[1]{\overline{\langle \; {#1}^{\vphantom{A}}
            \;\rangle}\;}
\def\bra#1{\langle \, {#1} \, | \;}
\def\ket#1{\; | \, {#1} \, \rangle}
\newcommand{\braket}[2]{\langle \, {#1} \, | \, {#2} \, \rangle}

\def\ketup{\ket{\uparrow}\;}

\newcommand{\pp}[2]{\frac{\partial \, {#1}}{\partial \, {#2}}\;}

\def\ppqmy#1{\frac{\partial \, {#1}}{\partial q_{\mu}}\;}

\newcommand{\ddt}{\frac{d}{dt}\;}

\newcommand{\vek}[1]{{\!\vec{\,#1}}}
\begin{document}
%
\typeout{   --- >>>   thermostat paper   <<<   ---   }
\typeout{   --- >>>   thermostat paper   <<<   ---   }
\typeout{   --- >>>   thermostat paper   <<<   ---   }
%
%
\journal{PHYSICA A}
\begin{frontmatter}
\title{Molecular dynamics investigations on a quantum system in a thermostat}
 
\author{J. Schnack\thanksref{JS}}
\address{Universit\"at Osnabr\"uck, Fachbereich Physik \\ 
         Barbarastr. 7, D-49069 Osnabr\"uck}

\thanks[JS]{email: juergen.schnack\char'100physik.uni-osnabrueck.de,\\
            WWW:~http://obelix.physik.uni-osnabrueck.de/$\sim$schnack}

\begin{abstract}

\noindent
The model quantum system of fermions in a one dimensional
harmonic oscillator potential is investigated by a molecular
dynamics method at constant temperature. Although in quantum
mechanics the equipartition theorem cannot be used like in the
Nos\'e--Hoover--thermostat it is possible to couple an additional
degree of freedom to the system which acts as a thermometer and
drives the system towards the desired temperature via complex
time steps.

\vspace{1ex}


\vspace{1ex}

\end{abstract}
\end{frontmatter}
\raggedbottom
\section{Introduction and summary}

Statistical properties of finite systems are of great
interest. The aim is to describe the behaviour of systems like
atomic clusters or atomic nuclei at finite temperatures and to
investigate properties like the specific heat or phase
transitions. These statistical properties are given by the
partition function which for classical systems in the canonical
ensemble reads
\begin{eqnarray}
\label{E-1-1}
Z
&=&
\int \prod_{i=1}^{N}\; \dint^3 x_i\; \dint^3 p_i\; 
\exp\left\{-\frac{1}{T} H\right\}
\ ,
\end{eqnarray}
where $H$ denotes the Hamilton function,
and for quantum systems
\begin{eqnarray}
\label{E-1-2}
Z
&=&
\tr\left(\exp\left\{-\frac{1}{T} \op{H}\right\}\right)
\ ,
\end{eqnarray}
where $\op{H}$ is the Hamilton operator (Throughout the article
operators are underlined by a tilde and $\hbar=c=k_B=1$.).
For realistic systems like atomic clusters or nuclei where the
Hamilton function or operator contains a (two--body)
interaction it is hard or impossible to evaluate the partition
function especially for the quantum description. 

Equations of motion for the investigated system are often much
easier; either they are exactly known and can be integrated at
least numerically as it is the case with the classical Hamilton's
equation or they can be approximated with standard methods like
TDHF or quantum molecular dynamics methods as it is the case on
the quantum side. The idea then is to extract the desired
thermodynamic quantities from the time evolution of the
system. If the system is ergodic, ensemble averages can be
replaced by time averages.

During the last decade a huge progress has been made on the
classical side of the problem (see for instance
\cite{Nos84,Hoo85,KBB90,Nos91}). 
To put it into a few words, the basic idea is to exploit the
equipartition theorem and use for instance the kinetic energy as
a measure of the current temperature. The system is cooled or
heated via pseudo--friction coefficients if the present
temperature is too high or too low, respectively. The equations
of motion are changed in the following way
\begin{eqnarray}
\label{E-1-3}
\ddt \vek{x}_k &=& \pp{H}{\vek{p}_k}
\ , \quad
\ddt \vek{p}_k = -\pp{H}{\vek{x}_k} - \vek{p}_k \; \zeta
\ , \qquad k = 1, \dots , N
\\[2mm]
\label{E-1-5}
\ddt \zeta &\propto& \left( E_{kin}(N) - \frac{3\,N}{2} T\right)
\ .
\end{eqnarray}
The last line, equation \fmref{E-1-5}, shows nicely how the
system is driven towards the temperature $T$ via the coupling to
the pseudo--friction coefficient $\zeta$. It was shown that the
resulting distributions are those of the canonical ensemble (see
e.g. the report \cite{Nos91}). In addition ergodicity can be
improved by using more pseudo--friction coefficients and
different couplings to the original system \cite{KBB90}.

In quantum mechanics the problem is much more involved. No
useful a priori relation between expectation values of
observables and temperature like in the equipartition theorem
can be exploited.  Attempts have been made to derive a thermal
non-linear Schr\"odinger equation which results in an ergodic
wave function \cite{Kus93}, but this method needs the relation
between mean energy and temperature as an input, which is not
known for most systems and therefore itself a matter of
investigation. A nontrivial quantum extension of the
Nos\'e--method has been derived in ref. \cite{GrT89} which will
be the subject of a forthcoming paper.

The idea of this article is to couple an additional degree of
freedom to the original system, which serves as a
thermometer. This idea was already successfully applied in
nuclear physics in order to determine the caloric curve of
nuclei and to investigate the nuclear liquid--gas phase
transition \cite{ScF97}.  The new aspect in this article is to
drive the system via complex time steps towards the desired
temperature.

The total system is described by a time--dependent state
$\ket{Q(t)}$ consisting of the original system $\ket{system(t)}$
and the thermometer $\ket{thermometer(t)}$
\begin{eqnarray}
\label{E-1-6}
\ket{Q(t)}
&=&
\ket{system(t)}\otimes\ket{thermometer(t)}
\ .
\end{eqnarray}
According to the difference between the temperature $T_{th}$
measured by the thermometer and the desired temperature $T$
the total system is evolved by a complex time step $d\tau$.
\begin{eqnarray}
\label{E-1-7}
d\tau = dt (1 -i d\beta) 
\ , \quad 
d\beta\propto (T_{th}-T)/T_{th}
\ , \quad 
\ket{Q(t)}
\rightarrow
\ket{Q(t+d\tau)}
\ .
\end{eqnarray}
As can be inferred from \fmref{E-1-7} $d\beta>0$ results in
cooling and $d\beta<0$ in heating of the system.

In the present article the proposed method will be examined with
the simple but basic quantum system of $N$ fermions in a common
one--dimensional harmonic oscillator potential. The advantage of
this system is that all relations of the canonical ensemble are
either known analytically or can easily be computed numerically.

In order to equilibrate the system in an ergodic sense a
short--range two--body repulsion among the particles is used
which does not destroy the implied picture of an ideal quantum
gas \cite{ScF96}. 

The results of time averaging are compared to the canonical
ensemble in terms of occupation numbers which in thermodynamic
equilibrium characterize the ideal Fermi gas enclosed in a
common harmonic oscillator potential completely.  It is
demonstrated that the system is indeed ergodic, i.e., the time
averaged occupations numbers coincide with those obtained in the
canonical ensemble.

\section{Method and setup}

\subsection{Model state and equations of motion}

In many cases the time--dependent Schr{\"o}dinger equation
cannot be solved. The following time--dependent quantum
variational principle (TDVP) \cite{KrS81}
\begin{eqnarray}
\label{E-2-1}
\delta \int_{t_1}^{t_2} \! \! dt \;
\bra{Q(t)}\; i \frac{d}{dt} - \op{H}\; \ket{Q(t)} \ &=&\ 0
\end{eqnarray}
allows to derive approximations to the time--dependent
Schr{\"o}dinger equation on the level of accuracy one needs or
can afford. For the variation of the trial state $\bra{Q(t)}$ in
the complete Hilbert space the TDVP leads to the Schr{\"o}dinger
equation. For trial states given in terms of parameters $Q(t) =
\{q_\nu(t)|
\nu=1,2,\dots\}$ the TDVP leads to Euler--Lagrange equations
which in their most general form can be written as
\begin{eqnarray}
\label{E-2-2}
\sum_\nu {\mathcal A}_{\mu \nu}(Q)\ \dot{q}_{\nu}
&=&
- \ppqmy{}{} \bra{Q(t)} \op{H} \ket{Q(t)}
\\
{\mathcal A}_{\mu\nu}(Q) 
&=&
\frac{\partial^{2}{\mathcal L}_{0}}{\partial \dot{q}_{\mu}
\partial q_{\nu}} -
\frac{\partial^{2}{\mathcal L}_{0}}{\partial \dot{q}_{\nu} 
\partial q_{\mu}}\ ,
\nonumber
\end{eqnarray}
where ${\mathcal L}_{0}=\bra{Q(t)}i \ddt\!\! \ket{Q(t)}$ and
${\mathcal A}_{\mu\nu}(Q)$ is a skew symmetric matrix.  The time
evolution of the parameters then defines the time dependence of
the many--body state $\ket{Q(t)}$.  If for instance $\ket{Q(t)}$
is chosen as a Slater determinant of arbitrary single--particle
states one ends up with time--dependent Hartree--Fock
\cite{KeK76}.

In this investigation the molecular dynamics approach of
Fermionic Molecular Dynamics (FMD) \cite{FMDRef} is taken
where $\ket{Q(t)}$ is a Slater determinant of single--particle
Gaussian wave--packets $\ket{q_l(t)}$
\begin{eqnarray}
\braket{\vec{x}}{q_l(t)} &=&
\exp\left\{ \, -\; \frac{(\,\vec{x}-\vec{b}_l(t)\,)^2}{2\,a_l(t)}
            \right\} 
\otimes\ketup
\; , \;
\vec{b}_l = \vec{r}_l + i a_l \vec{p}_l\ .
\label{E-2-3}
\end{eqnarray}
Each single--particle state is parametrized in terms of the
time--dependent mean position $\vec{r}_l(t)$, mean momentum
$\vec{p}_l(t)$ and the complex width $a_l(t)$. In the notation
of \fmref{E-2-3} the vector $\vec{b}_l$ is composed of
$\vec{r}_l$, $\vec{p}_l$ and $a_l$.  The time dependence of the
spin degrees of freedom is not considered, instead it is assumed
that all particles are identical fermions with the same spin
component$\ketup$. In order to perform complex time steps the
complex parameters $\vec{b}_l$ and $a_l$ are used.

The approach of FMD has several advantages, in connection with
this investigation it is important to realize that for the one--body
Hamiltonian
\begin{eqnarray}
\op{H} =
\sum_{l=1}^N\;\op{h}(l)\; , \quad
\op{h}(l) = \frac{\op{\vec{k}}^2(l)}{2 m} 
       + \half m \omega^2 \op{\vec{x}}^2(l)
\end{eqnarray}
of the harmonic oscillator the solution of the approximate
equations of motion \fmref{E-2-2} coincides with the exact
solution of the Schr{\"o}dinger equation. For fermions in a
harmonic oscillator these equations of motion can
be given explicitly \cite{ScF96} 
\begin{eqnarray}
\label{E-2-4}
{\displaystyle\ddt} \vec{b}_l = -i m \omega^2 a_l \vec{b}_l 
\; , \quad
{\displaystyle\ddt} a_l = - i m \omega^2 a_l^2 + 
                          \frac{\displaystyle i}{\displaystyle m}  
.
\end{eqnarray}
Replacing the complex $\vec{b}_l$ in eq. \fmref{E-2-4} yields
\begin{eqnarray}
\label{E-2-5}
\ddt \vec{r}_l = \frac{\vec{p}_l}{m}
\quad&\mbox{and}&\quad
\ddt \vec{p}_l = - m \omega^2 \vec{r}_l\ .
\end{eqnarray}
The parameters $\vec{r}_l$ and $\vec{p}_l$
follow the classical trajectories.  Nevertheless, the
parametrized trial state is the exact solution of the
Schr{\"o}din\-ger equation.  It is also worth to note that for
this example the equations of motion of the parameters are the
same, regardless whether the many--body state is
antisymmetrized, symmetrized or simply a product state.

\subsection{Time averaging}

The time averaging of an operator $\op{B}$ is done by taking the
trace with the statistical operator $\OpRerg$ \cite{ScF96}
\begin{eqnarray}
\label{E-2-6}
\OpRerg
&\; := \;&
\lim_{t_2\rightarrow\infty}\;
\frac{1}{(t_2 - t_1)}\;
\int_{t_1}^{t_2} \mbox{d}t\;
\ket{Q(t)}\bra{Q(t)}
\end{eqnarray}
which yields
\begin{eqnarray}
\label{E-2-7}
\TimeAv{\op{B}}
:= \;
\Tr\left(\OpRerg\;\op{B}\right)
=
\lim_{t_2\rightarrow\infty}\;
\frac{1}{(t_2 - t_1)}\;
\int_{t_1}^{t_2} \mbox{d}t\;
\bra{Q(t)}\op{B}\ket{Q(t)}\ .
\end{eqnarray}
In general the statistical operator $\OpRerg$ is a functional of
the initial state $\ket{Q(t_1)}$, the Hamilton operator
$\Operator{H}$ and the equations of motion.  If the ergodic
assumption is fulfilled, the statistical operator should only
depend on $\TimeAv{\op{H}}$.

\subsection{Thermometer and coupling}

In order to determine the temperature of the system it is
coupled to a thermometer. A thermometer is a device for which
thermodynamic relations, like those of kinetic energy and
temperature in the classical case, are known. Here the
thermometer is realized in terms of distinguishable particles in
another harmonic oscillator potential. 

\begin{figure}[hhhh]
\begin{center}
\epsfig{file=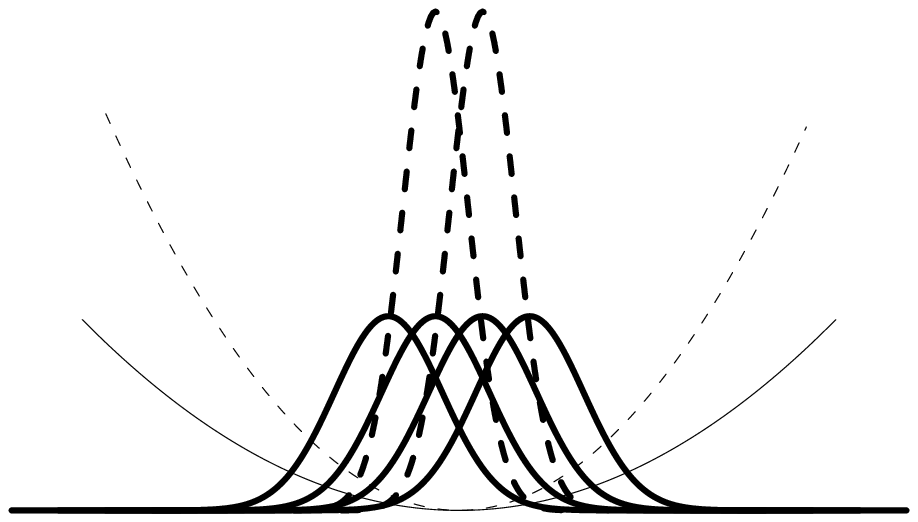,width=80mm}
\end{center}
\mycaption{Arrangement of fermion wave--packets in their
oscillator potential (solid lines) and thermometer wave--packets in
their oscillator (dashed lines).}{F-2-1}
\end{figure} 

Figure \xref{F-2-1} sketches the setup. The fermion system
consists of four Gaussian wave--packets (thick solid line)
enclosed in their harmonic oscillator potential (thin solid
line). The two thermometer wave--packets (thick dashed lines)
reside in the other harmonic oscillator (thin dashed line). The
complete system is represented by the state $\ket{Q(t)}$
\begin{eqnarray}
\ket{Q(t)}&=&\ket{Q_f(t)}\otimes\ket{Q_{th}(t)}
\nonumber
\\[2mm]
\label{E-2-8}
\ket{Q_f(t)}&=&\op{A}\left(
\ket{q_{f1}(t)}\otimes\ket{q_{f2}(t)}\otimes
\ket{q_{f3}(t)}\otimes\ket{q_{f4}(t)}
\right)
\\[2mm]
\ket{Q_{th}(t)}&=&\ket{q_{th1}(t)}\otimes\ket{q_{th2}(t)}
\nonumber
\end{eqnarray}
where $\ket{Q_f(t)}$ describes the fermions ($\op{A}$ being the
antisymmetrization operator) and $\ket{Q_{th}(t)}$, which is a
product state, the thermometer wave--packets.  The time
evolution according to \fmref{E-2-2} is driven by the Hamilton
operator
\begin{eqnarray}
\label{E-2-9}
\op{H}&=&\op{H}_f + \op{H}_{th} + \op{V}_{f,f} + \op{V}_{f,th}
\\
\op{H}_f&=&
\sum_i \left(
\frac{\op{k}_i^2}{2 m_f} + \half m_f \omega_f^2 \op{x}_i^2
\right)
\ , \quad
\op{H}_{th}=
\sum_i \left(
\frac{\op{k}_i^2}{2 m_{th}} + \half m_{th} \omega_{th}^2 \op{x}_i^2
\right)
\ ,
\nonumber
\end{eqnarray}
where $\op{H}_f$ is the Hamilton operator of the fermion
subsystem, $\op{H}_{th}$ the Hamilton operator of the
thermometer subsystem,
\begin{eqnarray}
\label{E-2-10}
\op{V}_{f,f}
&=&
\sum_{i<j} V_0 \exp{\left\{ -\frac{(\op{x}_i - \op{x}_j)^2}{r_0^2}\right\}}
\\
\omega_f &=& 8\MeV
\ , \quad
r_0 = \frac{0.01}{\sqrt{m_f \omega_f}}
\ , \quad
V_0 = (10^4\dots 10^5) \omega_f
\ ,
\nonumber
\end{eqnarray}
$\op{V}_{f,f}$ the short--range repulsion between all fermions
needed to equilibrate the fermion subsystem and $\op{V}_{f,th}$
the momentum dependent short--range repulsion between fermions
and thermometer wave packets, which is given in terms of its
matrix elements in the appendix.

There are several conditions that need to meet in order to let
the method work. The original system better has to be ergodic by
itself and the coupling of the thermometer must lead to thermal
equilibrium between the system and the thermometer.  These
conditions depend on the interaction which is needed to
equilibrate the system. Some experience \cite{ScF96,ScF97} and
physical reasoning show that such interactions should be short
ranged, repulsive and possibly momentum dependent. In addition
the time evolution has to run over many typical cycles of the
system so that enough collisions happen.  I also found that the
thermometer works best if $\omega_{th} > \omega_f$ so that the
thermometer has a smaller heat capacity and that the mass
$m_{th} < m_f$ so that the thermometer wave packets cross the
region of the fermions. I chose $\omega_{th} = \pi \omega_f$ and
$m_{th} = m_f/\pi^2$.

If the system is ergodic the total energy of the thermometer
subsystem determines the temperature $T_{th}$
\begin{eqnarray}
\label{E-2-11}
T_{th}
&=&
\omega_{th}
\left[
\ln\left(
\frac{E_\Th + N \frac{\omega_{th}}{2}}{E_\Th 
- N \frac{\omega_{th}}{2}}
\right)
\right]^{-1}
\ .
\end{eqnarray}

\section{Results}

\subsection{Canonical ensemble of fermions in a harmonic oscillator}

Since all results of the time averaging are compared to the
canonical ensemble, the canonical ensemble is shortly explained
for $N$ fermions in a common one--dimensional harmonic
oscillator potential $\op{H}_f$. The statistical operator is
given by
\begin{eqnarray}
\label{E-3-1}
\Operator{R}(T)
&=&
\frac{1}{Z(T)}
\exp\left\{-\frac{\op{H}_f}{T}\right\}
\\
\op{H}_f
&=&
\sum_{l=1}^N
\Operator{h}(l)
\ ,\qquad
\Operator{h}(l)
=
\omega_f \; \sum_{n=0}^\infty 
\Big(n + \half\Big)\; \Operator{c}_n^+\Operator{c}_n\ ,
\nonumber
\end{eqnarray}
then the statistical mean of an operator $\Operator{B}$ 
is calculated as
\begin{eqnarray}
\label{E-3-2}
&&\hspace{-1mm}
\EnsembleMean{\Operator{B}} 
:=
\Tr\left(\Operator{R}(T)\;\Operator{B}\right)
\end{eqnarray}
In eq. \fmref{E-3-2} the subscript $T$ indicates that the
average in the canonical ensemble is taken at a constant
temperature $T$.

\begin{figure}[!bt]
\begin{center}
\unitlength1mm
\begin{picture}(140,44)
\put( 0,0){\epsfig{file=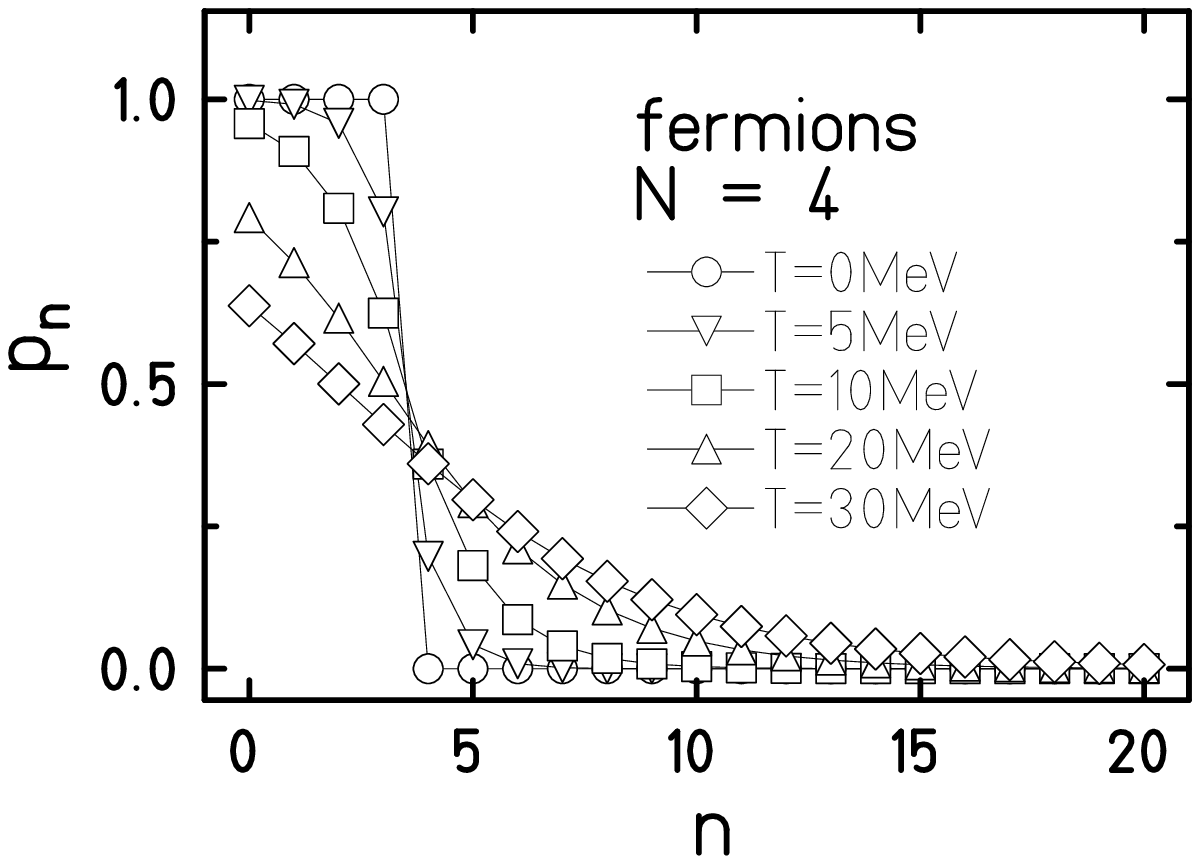,width=68mm}}
\put(75,0){\epsfig{file=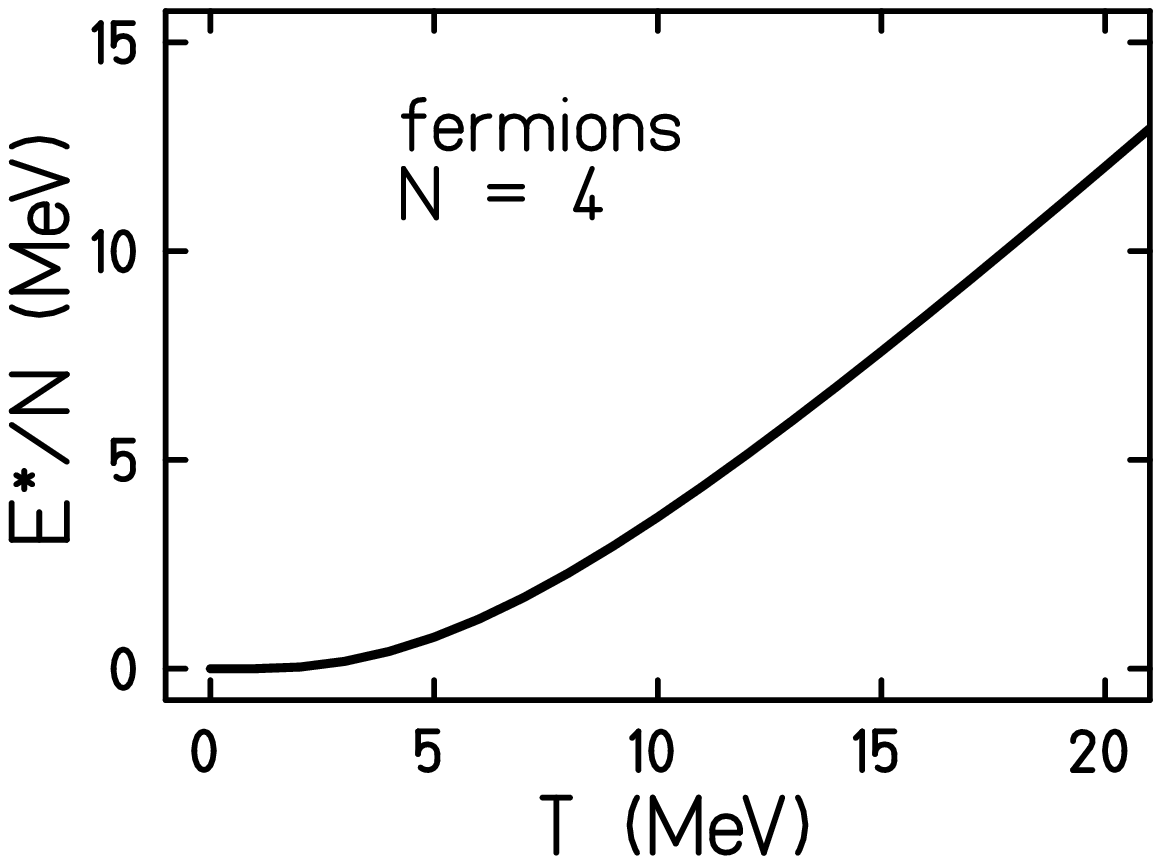,width=65mm}}
\end{picture}
\end{center}
\mycaption{Canonical ensemble for four fermions:
        Occupation numbers $p_n$ of the oscillator eigenstates
	for five temperatures (l.h.s.) and excitation energy as
	a function of temperature (r.h.s).}{F-3-1}
\end{figure} 

The most prominent physical quantities which will be
investigated in the following are the mean energy, which can be
represented in closed form as
\begin{eqnarray}
\label{E-3-3}
\EnsembleMean{\op{H}_f} 
&=&
E_0(N) + 
\sum_{k=1}^{N} k \frac{\omega_f}{2}
\left[ \mbox{coth}\Bigg(k \frac{\omega_f}{2\, T}\Bigg) -1
\right]
\\[2mm]
&&E_0(N) = N^2 \frac{\omega_f}{2}
\nonumber
\end{eqnarray}
and the mean occupation probabilities $p_n$
\begin{eqnarray}
\label{E-3-4}
p_n = \EnsembleMean{\Operator{c}_n^+\Operator{c}_n}\ ,
\end{eqnarray}
where $\Operator{c}_n^+$ denotes the creation operator of a
fermion in the single--particle energy--eigenstate $\ket{n}$.
The occupation probabilities are displayed in \figref{F-3-1} on
the l.h.s. for various temperatures, the r.h.s. shows the
dependence of the mean excitation energy on the temperature.

\subsection{Time evolution in the thermostat}

The deterministic time evolution of the system is followed over
about 400 cycles ($\equiv 60000$fm/c) of the fermion
system. Initially the four--fermion system is far from
equilibrium which can be inferred from
\figref{F-3-2} (l.h.s.) where the occupation probabilities of
an initial state $\ket{Q_f(t=0)}$ are displayed.

\begin{figure}[!bt]
\begin{center}
\unitlength1mm
\begin{picture}(140,31)
\put(  0,0){\epsfig{file=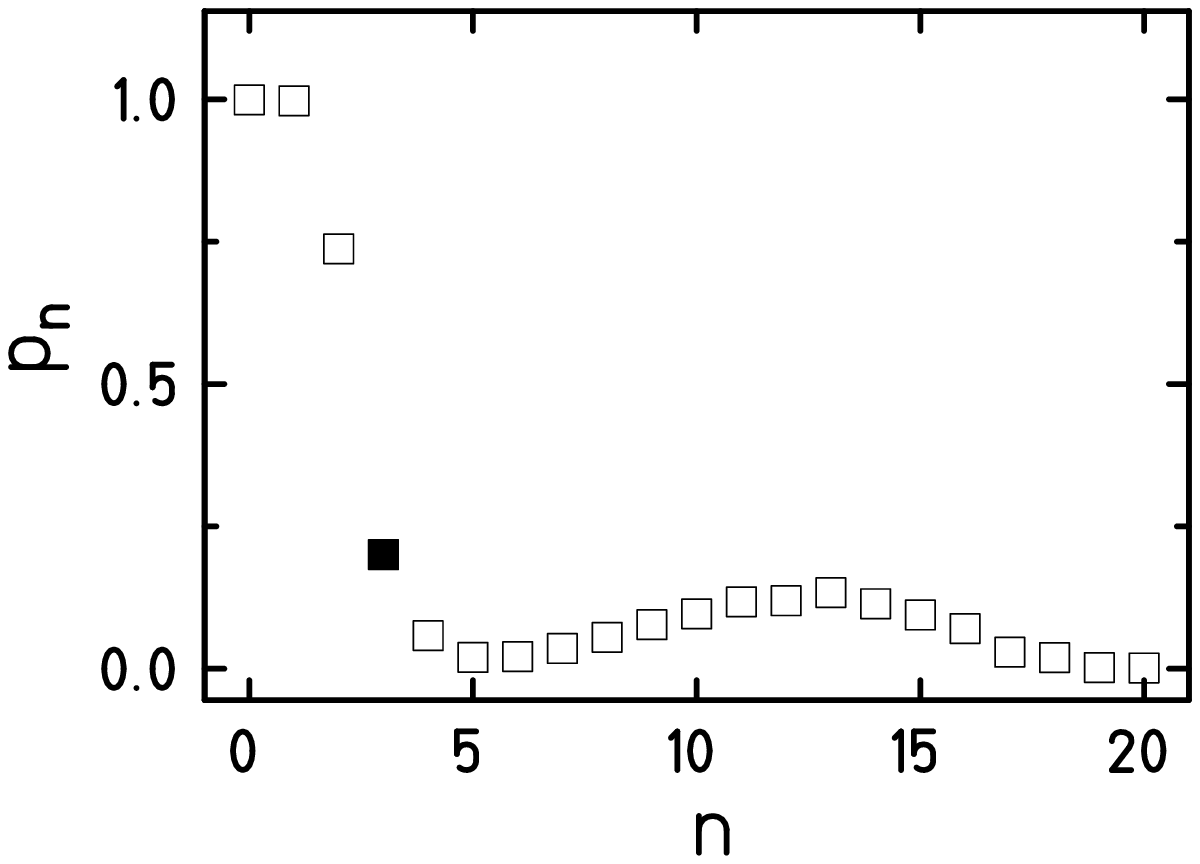,width=45mm}}
\put( 47,0){\epsfig{file=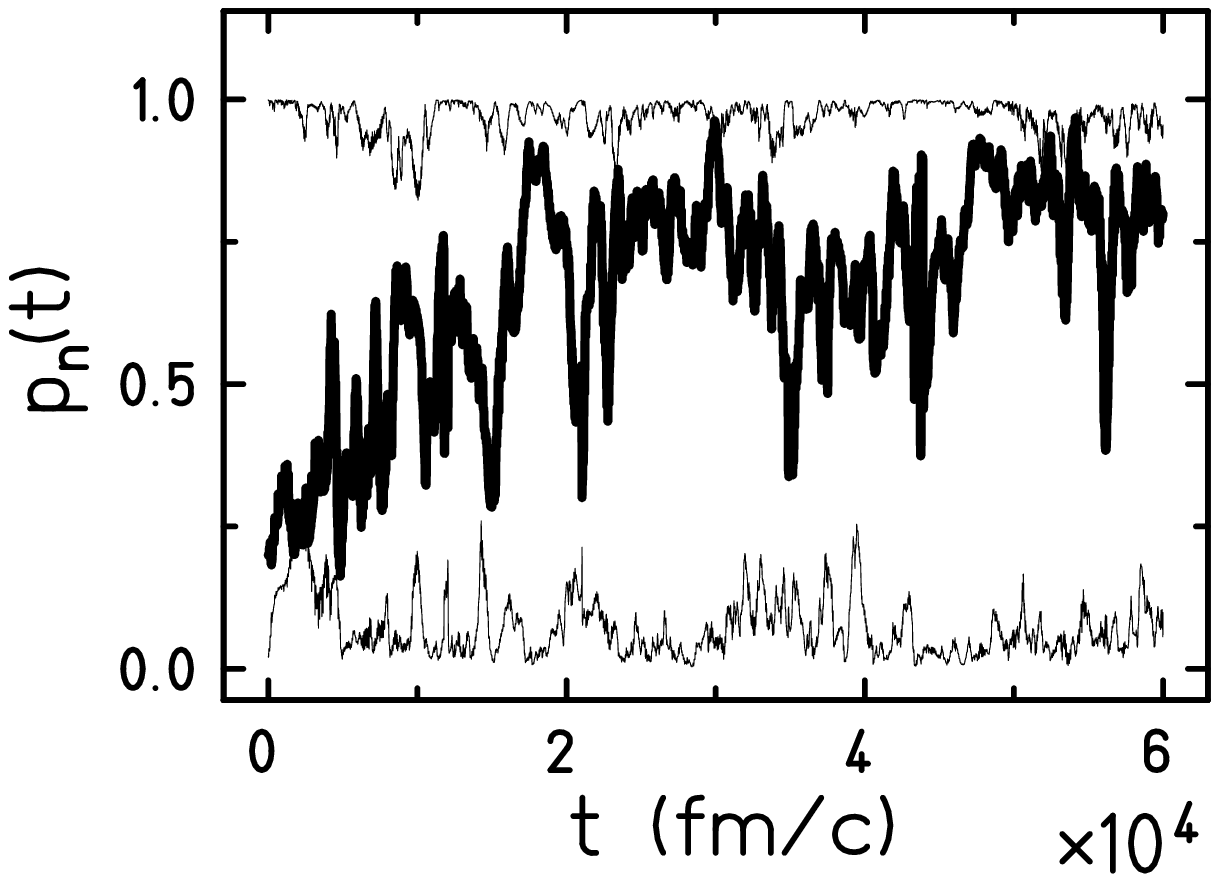, width=45mm}}
\put( 95,0){\epsfig{file=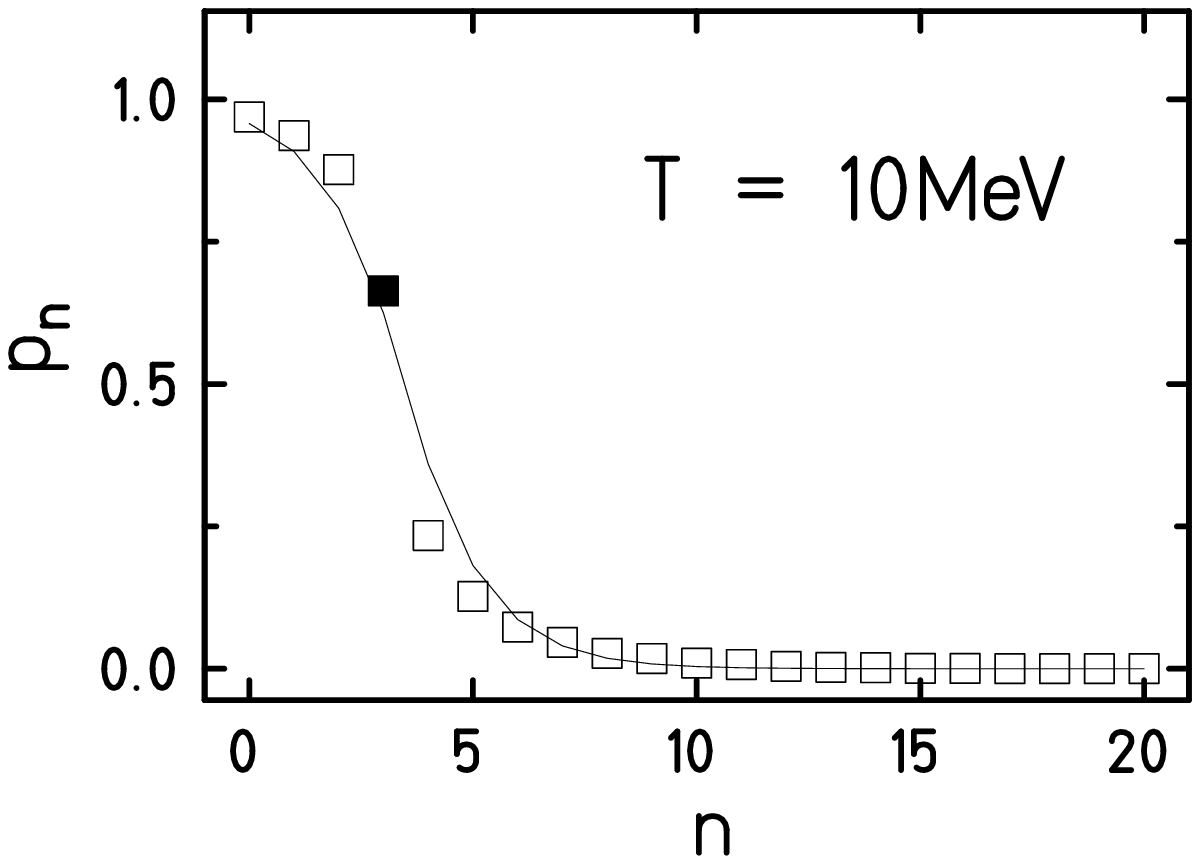, width=45mm}}
\end{picture}
\end{center}
\mycaption{Occupation probabilities $p_n$; l.h.s.: distribution
at $t=0$; middle: time evolution $p_n(t)$ for $n=0$ --- top
line, $n=3$ --- middle, $n=6$ --- bottom line; r.h.s.: time
averaged distribution (symbols), the line shows the canonical
result.}{F-3-2}
\end{figure} 

Due to the interaction between all wave packets and due to the
heating or cooling according to the present temperature the
occupation numbers are re-shuffled with time. This is seen in
\figref{F-3-2} (middle) where the time evolution of the momentary
occupation numbers $p_n(t)$ is presented for $n=0$, $n=3$ and
$n=6$. The picture reminds of deterministic chaos. The result of
time averaging is shown on the r.h.s. of \figref{F-3-2}
(symbols) where the dependence obtained in the canonical
ensemble is given by the solid line (see also \figref{F-3-1}).
How the system evolves from non-equilibrium towards equilibrium
can be visualized nicely by the occupation number of state 3
which in all three figures is given in bold.

\begin{figure}[hhhh]
\begin{center}
\epsfig{file=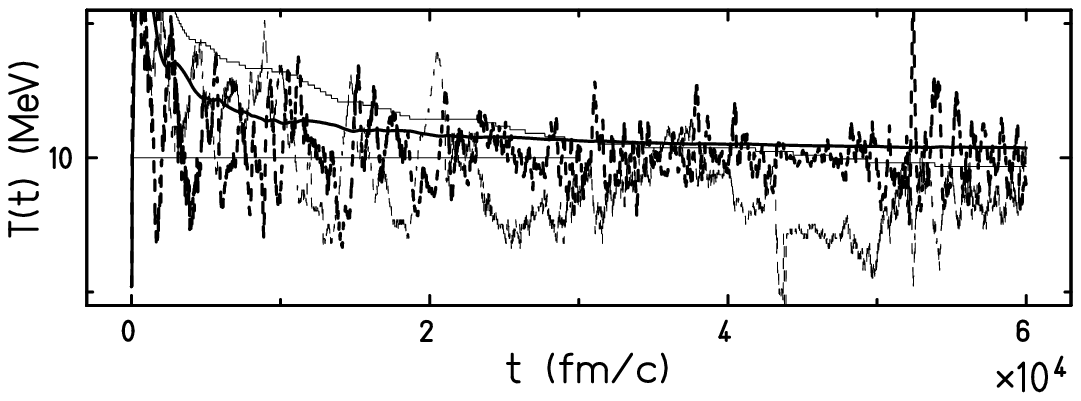,width=120mm}
\end{center}
\mycaption{Time evolution of actual temperatures (dashed lines)
and time averaged temperatures (solid lines) for the thermometer
system (thick lines) and for the fermion system (thin
lines).}{F-3-3}
\end{figure} 

Since the relation between temperature and mean energy is known
for the fermion system (see eq. \xref{E-3-3}), it can be checked
how well the fermion temperature is reflected by the
thermometer. Figure \xref{F-3-3} shows four temperatures, two
for the thermometer (thick lines) and two for the fermion system
(thin lines). The solid lines are related to the time averaged
energies of the fermion and thermometer subsystems,
respectively, where the averaging is performed from the
beginning until the current time. The dashed lines show the
momentary temperatures.  One sees that the fermion system is quite
excited initially whereas the thermometer starts being cold.
One also realizes that the momentary temperatures fluctuate
strongly, which is not astonishing since both subsystems are
very small. Taking $1/\sqrt{N}$ as a rough estimate for relative
fluctuations it becomes clear that for instance in the fermion
subsystem ($N=4$) fluctuations of 50\% of the mean value are
quite normal.

\begin{figure}[hhhh]
\begin{center}
\epsfig{file=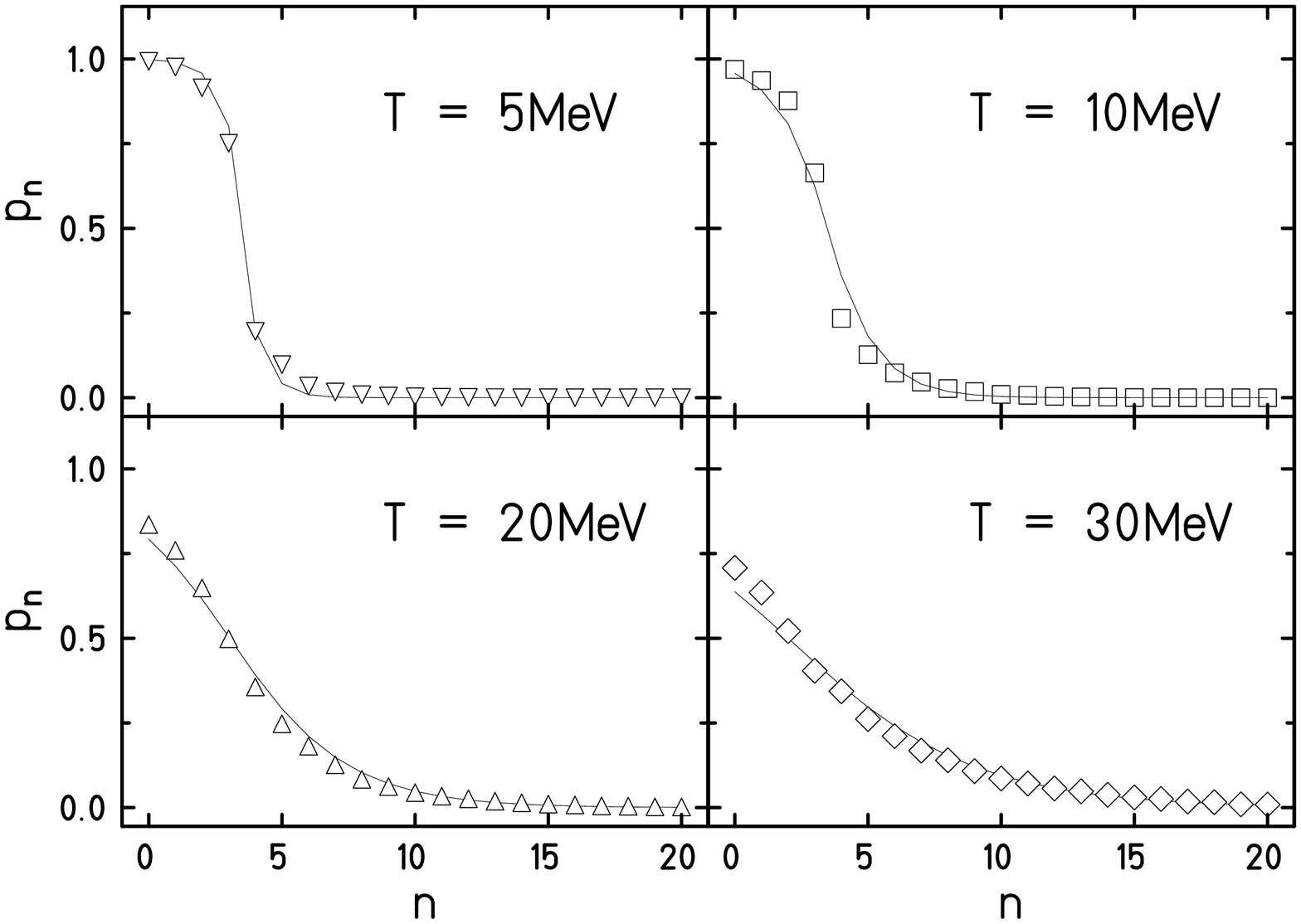,width=120mm}
\end{center}
\mycaption{Time averaged occupation numbers (symbols) compared
with those of the canonical ensemble (solid lines).}{F-3-4}
\end{figure} 

The simulation explained in detail for $T=10\MeV$ is repeated
for other temperatures. The resulting occupation probabilities
are displayed in \figref{F-3-4}. It is impressive that the time
averaged occupation numbers match those of the canonical
ensemble for a wide range of temperatures almost exactly,
\begin{eqnarray}
\label{E-3-5}
\TimeAv{\Operator{c}_n^+\Operator{c}_n}
\approx
\EnsembleMean{\Operator{c}_n^+\Operator{c}_n}
\qquad \forall n
\ ,
\end{eqnarray}
meaning that the system is ergodic, i.e., that time averages
approach ensemble averages of the canonical ensemble.

Nevertheless, the proposed method has also limitations. The
necessary interaction of all constituents prevents accurate
investigations close to the ground state, because the system is
alway excited due to the interaction. For a quantum system close
to the ground state this leads already to a large temperature
deviation (see \figref{F-3-2}). Methods to circumvent this
problems are under investigation.

\appendix
\section{Interaction of fermions and thermometer wave packets}

The interaction between fermions (index $k$) and thermometer
wave packets (index $l$) is modeled via the following expression
\begin{eqnarray}
\label{E-A-1}
\bra{Q} \op{V}_{f,th} \ket{Q}
&=&
\sum_{k,l} \frac{V_0}{R_{kl}}\;
(p_k - p_l)^2
\exp\left\{ - \frac{(r_k - r_l)^2}{R_{kl}^2}   \right\} 
\nonumber \\[2mm]
R_{kl}^2 = r_0^2 + 
0.2\left(
\frac{|a_k|^2}{a_{kR}} + \frac{|a_l|^2}{a_{lR}}
\right)
&,&
r_0 = \frac{0.01}{\sqrt{m_f \omega_f}}
\ , \quad
V_0 = (10^{-1}\fm^{3}\dots 1\fm^{3})\; \omega_f
\ .
\nonumber
\end{eqnarray}

{\bf Acknowledgments}\\[5mm] I would like to thank Prof. Klaus
B\"arwinkel and Christian Schr\"oder for reading the manuscript.

\end{document}